# Enhanced ferromagnetism of CrI$_3$ bilayer by self-intercalation

Yu Guo, Nanshu Liu, Yanyan Zhao, Xue Jiang, Si Zhou,[*] Jijun Zhao

*Key Laboratory of Materials Modification by Laser, Ion and Electron Beams, Dalian University of Technology, Ministry of Education, Dalian 116024, China*

**Abstract**

Two-dimensional (2D) ferromagnets with high Curie temperature have long been the pursuit for electronic and spintronic applications. CrI$_3$ is a rising star of intrinsic 2D ferromagnets, however, it suffers from weak exchange coupling. Here we propose a general strategy of self-intercalation to achieve enhanced ferromagnetism in bilayer CrI$_3$. We showed that filling either Cr or I atoms into the van der Waals gap of stacked and twisted CrI$_3$ bilayers can induce the double exchange effect and significantly strengthen the interlayer ferromagnetic coupling. According to our first-principles calculations, the intercalated native atoms act as covalent bridge between two CrI$_3$ layers and lead to discrepant oxidation states for the Cr atoms. These theoretical results offer a facile route to achieve high-Curie-temperature 2D magnets for device implementation.

[*] Corresponding author. Email: sizhou@dlut.edu.cn

The discoveries of magnetism at the atomic layer limit have opened a new avenue for the research of two-dimensional (2D) materials.[1-3] Recently, monolayer $CrI_3$ has been synthesized in the experiment[4] and confirmed to be an intrinsic 2D Ising ferromagnetic (FM) material.[5] It possesses many excellent characters, such as perfect crystalline order, strong anisotropy in magnetic configuration, and a suitable band gap of about 1.2 eV.[6] However, the Curie temperature of monolayer $CrI_3$ is only 45 K due to the weak intralayer exchange,[7] while bilayer $CrI_3$ with monoclinic stacking even exhibits antiferromagnetic (AFM) spin order between the two layers, which hinders the practical application of this novel 2D material.[7] Therefore, it is highly desirable to seek effective strategies to enhance the ferromagnetism and raise the Curie temperature of 2D $CrI_3$.

Intercalation has also been demonstrated as a powerful approach to modulate the physical and chemical properties of 2D layered materials. So far, various atoms, ions, or molecules can be intercalated in numerous kinds of 2D materials with well controlled concentration, triggering exotic phenomena for practical applications. For instance, Li intercalation of graphene film remarkably increases optical transmittance and electrical conductivity,[8] while Ca intercalation of bilayer graphene drives superconductivity.[9] 2D $SnS_2$ intercalated by different transition metal atoms can exhibit $p$-type, $n$-type and degenerately doped semiconducting behavior.[10] High Curie temperature up to ~300 K can be achieved in 2D magnets $Fe_{3-x}GeTe_2$ and $Cr_2Ge_2Te_6$ by intercalation of Na and organic ions, respectively.[11-12] $TaS_2$ sheets intercalated with Fe atoms using chemical vapor transport method show tunable magnetic order, magnetic anisotropy, and magnetoresistance.[13-15] Cl intercalation of g-$C_3N_4$ monolayer promotes charge carrier migration and yields favorable band gap and band edge positions to improve the photocatalytic performance for water splitting and $CO_2$ reduction.[16] Furthermore, intercalation of native atoms into bilayer transition metal dichalcogenides during growth is feasible in laboratory, which allows the generation of a new class of covalently bonded materials with stoichiometry- or composition-dependent properties. For example, by controlling the metal chemical potential, a range of self-intercalated compounds of $TaS(Se)_2$ can be obtained, such as Ta$_9$S$_{16}$, Ta$_7$S$_{12}$, Ta$_{10}$S$_{16}$, Ta$_8$Se$_{12}$ and Ta$_9$Se$_{12}$, some of which exhibit ferromagnetism or the unique Kagome lattice.[17]

Motivated by recent experimental advances in the intercalated 2D materials, herein we propose a feasible strategy to elevate the Curie temperature for stacked and twisted bilayer CrI$_3$ by self-intercalation of Cr or I atoms. By first-principles calculations, the geometries, electronic band structures, and magnetic behavior of bilayer CrI$_3$ with various intercalation configurations and concentrations have been systematically investigated. The key parameters that govern the exchange energy of intercalated bilayer CrI$_3$ have been determined, and the underlying double exchange mechanism was elucidated.

First-principles calculations were performed using spin-polarized density functional theory in conjunction with the projector augmented wave potentials, as implemented in the VASP.[18-19] To describe exchange-correlation interactions, we used the generalized gradient approximation with the Perdew-Burke-Ernzerhof functional.[20] The energy cutoff for the planewave basis was chosen as 500 eV. The convergence criteria of total energy and residual force on each atom were set to be $10^{-7}$ eV and 0.01 eV/Å, respectively. A vacuum region of 20 Å was added to the perpendicular direction to eliminate the interaction between periodic images. Uniform **k**-point meshes with spacing of ~0.015 Å$^{-1}$ were adopted to sample the 2D Brillouin zones.

We started from the crystal structures of bilayer CrI$_3$ with different stacking geometries. Two types of stacked bilayer CrI$_3$ were considered, namely low-temperature (LT) and high-temperature (HT) phases, as shown in Fig. 1a and 1b, respectively. They have nearly identical lattice parameter (6.97 Å) and interlayer distance (3.54 Å). In order to explore the preferred magnetic interaction, we defined the exchange energy per formula (f.u.) as

$$\Delta E = (E_{\text{AFM}} - E_{\text{FM}})/n \qquad (1)$$

where $E_{\text{AFM}}$ and $E_{\text{FM}}$ are the energies of CrI$_3$ bilayer with AFM and FM spin configurations between the two layers, respectively; $n$ is the number of CrI$_3$ formula in the supercell. Positive (negative) value of $\Delta E$ represents FM (AFM) spin order between the two CrI₃ layers. Larger $\Delta E$ means stronger interlayer FM coupling and generally leads to higher Curie temperature.[21] As given in Table 1, $\Delta E$ for pristine bilayer CrI₃ are as small as 2.19 meV/f.u. (LT) and −0.14 meV/f.u. (HT), respectively, which are consistent with previous experimental observation of weak ferromagnetism in LT phase and antiferromagnetism in HT phase and close to the reported theoretical values (3.90 meV/f.u. for LT phase and –0.17 meV/f.u. for HT phase).[4, 7, 21-24]

During the growth process, 2D layered materials with twisted angles are commonly observed in experiments,[25-27] offering extra degree of freedom for modulating their physical properties.[28-32] Therefore, we also considered bilayer CrI₃ with twisted angles of 21.79° and 38.42° displayed in Fig. 1 and Fig. S1 in the Supplementary Material (SM), thereafter named as 21.79°-CrI₃ and 38.42°-CrI₃, respectively. The interlayer binding energies of these twisted CrI₃ bilayers are nearly identical to that of the LT and HT phases (0.39 meV/atom), indicating favorable formation of both twisted and stacked CrI₃ sheets in the experiment. The considered twisted systems exhibit interlayer ferromagnetic coupling with smaller exchange energies than that of the LT phase, i.e. $\Delta E$ = 0.61 and 0.99 meV/f.u. for 21.79°-CrI₃ and 38.42°-CrI₃, respectively. Apparently, it is difficult to enhance the ferromagnetism of bilayer CrI₃ simply by twisting the two layers, and other strategies are desired to increase the exchange energy or even change the exchange mechanism.

Inspired by successful intercalation in various layered materials,[11-15, 17, 33-41] here we explore the possibility of self-intercalation of Cr or I atoms into bilayer CrI₃, as displayed in Fig. 1 and Fig. S1 in the SM. Starting from $\sqrt{3}\times\sqrt{3}$, 2 × 2, $\sqrt{7}\times\sqrt{7}$ and 3 × 3 supercells of the LT and HT phases, a Cr or I atom is intercalated between CrI₃ layers, resulting in different concentrations (defined as the ratio of one Cr or I atom to the number of CrI₃ formula in the supercell). For all the systems, the intercalated Cr or I atom is always bonded with the nearby I atoms with bond lengths of around 2.79 and 2.92 Å, respectively, which are close to the intralayer Cr−I bond length of 2.77 Å. Compared with the interlayer spacing of 3.54 Å for the pristine bilayer, the intercalated Cr atom stays on the hollow site and reduces the interlayer distance to about 3.40 Å, while the intercalated I atom locates on the bridge site of the two nearest intralayer I atoms and enlarges the interlayer spacing to up to 4.26 Å. Such difference originates from the distinct Cr−I and I−I bond nature, which is also reflected in the intercalation energies for Cr and I atoms discussed below.

To characterize the energetic stability of intercalated atoms, we define the intercalation energy as

$$E_{int} = E_{tot} - E_{bilayer} - E_{Cr/I} \qquad (2)$$

where $E_{tot}$ and $E_{bilayer}$ are the energies of bilayer $CrI_3$ with and without intercalation, respectively; $E_{Cr/I}$ is the energy of a single Cr or I atom. As listed in Table S1 of SM, intercalation energies of Cr (I) atom for the LT and HT phases are in the range of −3.97 ~ −4.11 eV (−0.13 ~ −0.31 eV) and −4.29 ~ −4.39 eV (−0.69 ~ −1.61 eV), respectively. The twisted systems have $E_{int}$ close to the values of the LT phase (about −4.00 and −0.39 eV for Cr and I intercalation, respectively). These results suggest that self-intercalation of Cr or I atoms into the vdW gap of stacked or twisted $CrI_3$ bilayer is energetically feasible. Compared with I intercalation, the magnitude of $E_{int}$ for the Cr-intercalated systems is several times larger, again indicating the stronger bond formation between the intercalated Cr atom and nearest in-plane I atoms. Interestingly, the magnitudes of these intercalation energies are comparable to the formation energies of single Cr and I vacancy in the pristine $CrI_3$ layers (4.00 and 1.15 eV, respectively), implying that the escaped Cr and I atoms from vacancy defects may be self-intercalated into the vdW gap between $CrI_3$ layers. Moreover, tearing the intercalated bilayer off bulk $CrI_3$ involves an exfoliation energy of 0.20 J/m$^2$, almost the same as that for ripping monolayer $CrI_3$ from the bulk material (0.21 J/m$^2$, see Fig. S4 in the SM for details). Thus, self-intercalation would not affect the exfoliation behavior of layered $CrI_3$. The thermal stability of intercalated $CrI_3$ bilayers was further assessed by the Born-Oppenheimer molecular dynamics (BOMD) simulation (see Fig. S2 in the SM). The intercalated $CrI_3$ bilayers can maintain their structures at 300 K for at least 10 ps simulation with the largest deviation of 0.41 Å, demonstrating their good stability at room temperature. Moreover, the diffusion behavior of the intercalated Cr atom in the vdW gap of LT and HT phases was examined (Fig. S3 in the SM), which involves moderate barriers of 0.54 and 0.74 eV, respectively, and implies the easy migration of Cr atom to fill in the vdW gap.

Most encouragingly, self-intercalating Cr or I atoms into bilayer $CrI_3$ can significantly increase the exchange energy ($\Delta E$) and elevate the Curie temperature. All the considered stacking and twisted systems become ferromagnetic order for the interlayer spin configuration. As displayed by the spin charge density in Fig. S5 in the SM, the magnetic moments are carried by both intralayer and intercalated Cr atoms. Table 1 presents $\Delta E$ values in the range of 6.78 ~ 38.58 meV/f.u. for Cr intercalation and 3.15 ~ 21.05 meV/f.u. for I intercalation, which is one or two orders of magnitude larger than the values of pristine $CrI_3$ bilayers (2.95 meV/f.u. for LT phase, −0.14 meV/f.u. for HT phase, 0.61 meV/f.u. for 21.79°-$CrI_3$, and 0.99 meV/f.u. for 38.42°-$CrI_3$). Especially, the exchange energy increases linearly with the intercalated Cr concentration (Fig. 2), suggesting that the magnetism of $CrI_3$ bilayer can be effectively modulated by Cr intercalation. According to our previous study, the exchange energy of $CrI_3$ bilayer can be increased up to 9.42 meV/f.u. by proximity effect, which elevates $T_c$ up to 130 K.[42-43] The exchange energy is positively correlated with Curie temperature and can reflect the trend of robustness of FM order. The present self-intercalated $CrI_3$ bilayers with much larger exchange energies would possess even higher Curie temperature than 130 K. Similar results are also found for trilayer and bulk $CrI_3$ by Cr intercalation with exchange energies increased to 19.66 meV/f.u. for the LT phase (3.30 meV/f.u. for pristine systems), while AFM-to-FM transition occurs for the HT phase (see Fig. S8 and Table S3 in the SM). Therefore, self-intercalation is an effective strategy to enhance the ferromagnetism of layered $CrI_3$.

The enhanced interlayer FM coupling can be understood by the decomposed charge density, electronic band structures and density of states in Fig. 3, Fig. S5, Fig. S6 and Fig. S7 in the SM. Under octahedral crystal field, the fivefold degenerate $d$ orbitals of Cr atom split into a threefold occupied $t_{2g}$ ($d_{xy}$, $d_{x^2-y^2}$ and $d_{z^2}$) orbital and a twofold unoccupied $e_g$ ($d_{xz}$ and $d_{yz}$) orbital. Self-intercalation induces notable electronic states in the gap that promote the hybridization between $e_g$ and $t_{2g}$ states, which in turn is the key for strengthening the FM state of bilayer CrI$_3$. The decomposed spin-up charge density of conduction band minimum further reveals the gradually increasing occupation of $e_g$ orbital as the concentration of intercalated Cr atom increases. The $e_g$–$t_{2g}$ coupling strength can be quantitatively characterized by the virtual exchange gap ($G_{ex}$) between the $e_g$ and $t_{2g}$ orbitals,[44] which can be measured from the projected density of states. Taking Cr intercalation as a representative, $G_{ex}$ values for various intercalated bilayers are in the range of 0.37 ~ 0.83 eV, remarkably smaller than the values of LT (0.98 eV) and HT (0.92 eV) phases. Generally speaking, $G_{ex}$ decreases with increasing Cr concentration, leading to the gradually increased $\Delta E$ and enhanced ferromagnetism of bilayer CrI$_3$ (see Fig. 2).

To gain further insights into the impact of self-intercalation on the electronic band structure of bilayer CrI$_3$, we examined the bond nature, charge transfer and oxidation states. As revealed by the electron density counter plot in Fig. 4a, the two CrI$_3$ layers have interlayer bonding through the intercalated Cr or I atoms. The intercalated Cr atom gains 0.02 ~ 0.12 $e$ from the intralayer Cr atoms nearby according to Bader charge analysis in Table S5 in the SM, thereby modifying the oxidation states of Cr atoms. Taking the HT phase as a representative (see Fig. 4b), the intercalated Cr atom has a formal valence state of +2, while the two nearest intralayer Cr atoms (sharing an I atom with intercalated Cr) have a formal valence state of +2.67, compared with +3 for Cr in pristine CrI$_3$ bilayer. The presence of intercalated and intralayer Cr atoms with different oxidation states signifies the double exchange mechanism[45-46] that dominates the interlayer ferromagnetism in CrI$_3$ bilayer, as illustrated by Fig. 4c. In the Cr$^{\delta+}$–Cr$^{2+}$ (2 < $\delta$ < 3) mixed valence complexes, an electron hops between Cr$^{\delta+}$ and Cr$^{3+}$ cations through a bridging I$^-$ anion. Within the double exchange picture, electron transfer from one Cr cation to another would be more easily if it does not have to flip spin orientation in order to conform with Hund's rule, which thus facilitates the ferromagnetic coupling between two Cr cations with different valences. Similar double exchange mechanism is also applicable to bilayer CrI$_3$ with I intercalation (see Fig. S9). Even though the intercalated I atom is not spin-polarized, it donates certain amounts of electrons (0.06

~ 0.11 *e*) to the adjacent I atoms in the two CrI$_3$ layers, which in turn leads to different valence states for the nearby intralayer Cr atoms.

In summary, we significantly enhance the interlayer ferromagnetic order of bilayer CrI$_3$ by self-intercalation of Cr or I atoms. The magnetic behavior of intercalated bilayer CrI$_3$ of the LT and HT stacking geometries or with a twisted angle has been systematically explored as a function of intercalated concentration. Both Cr and I intercalation lead to robust interlayer ferromagnetic order for all the considered CrI$_3$ bilayers. The exchange energy increases with the intercalated Cr concentration, reaching about 40 meV/f.u. compared to the values of pristine CrI$_3$ bilayer (2.95 meV/f.u. for LT phase and −0.14 meV/f.u. for HT phase). Such strong ferromagnetism is dominated by the double exchange mechanism originated from the interlayer charge transfer via the intercalated atoms, which results in different oxidation states for the Cr atoms in CrI$_3$ bilayers and enhanced $e_g$-$t_{2g}$ hybridization. These theoretical results provide a universal and experimentally feasible strategy to effectively raise the Curie temperature of 2D ferromagnets via self-intercalation for practical uses.

**Acknowledgement**

This work was supported by the China Postdoctoral Science Foundation (BX20190052, 2020M670739), the National Natural Science Foundation of China (11974068), and the Fundamental Research Funds for the Central Universities of China (DUT20LAB110). The authors acknowledge the computer resources provided by the Supercomputing Center of Dalian University of Technology and Shanghai Supercomputer Center.

Table 1. Exchange energy ΔE (in the unit of meV/f.u.) of pristine, Cr and I intercalated CrI$_3$ bilayers, including HT and LT phases, twisted bilayers 21.79°-CrI$_3$ and 38.42°-CrI$_3$. The numbers in the brackets are the exchange energy for 21.79°-CrI$_3$ and 38.42°-CrI$_3$ without intercalation.

| | ΔE | pristine | √3×√3 | 2×2 | √7×√7 | 3×3 | 21.79° | 38.42° |
|---|---|---|---|---|---|---|---|---|
| Cr intercalation | LT | 2.95 | 38.58 | 31.48 | 21.33 | 6.78 | 5.38 (0.61) | 13.17 (0.99) |
| | HT | −0.14 | 29.99 | 28.50 | 18.18 | 15.02 | | |
| I intercalation | LT | 2.95 | 6.64 | 6.44 | 3.15 | 18.64 | 3.67 | 6.30 |
| | HT | −0.14 | 21.05 | 16.50 | 6.31 | 18.23 | | |

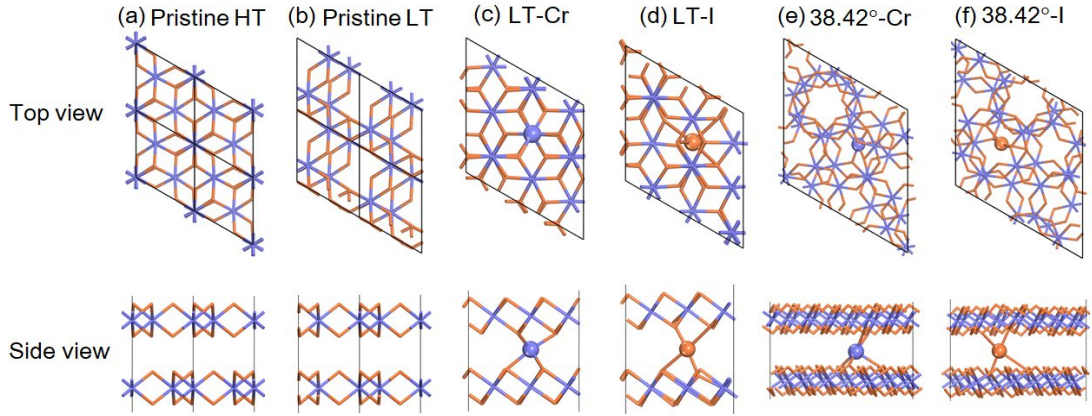

Fig. 1. Atomic structures of pristine (a) LT and (b) HT phases of bilayer CrI$_3$, (c) Cr and (d) I intercalated LT phase of $\sqrt{3}\times\sqrt{3}$ supercell, and (e) Cr and (f) I intercalated twisted 38.42°-CrI$_3$. The Cr and I atoms are shown in purple and orange colors, respectively.

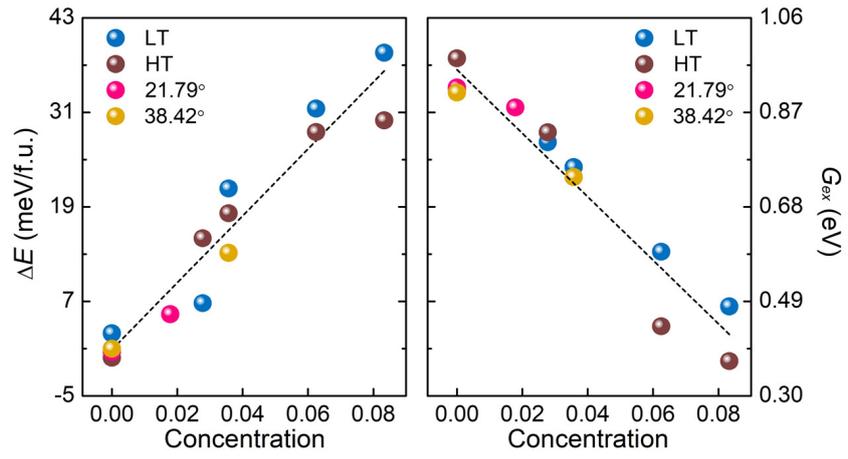

Fig. 2. Exchange energy (ΔE) and virtual exchange gap ($G_{ex}$) of various Cr-intercalated $CrI_3$ bilayers as a function of Cr intercalation concentration.

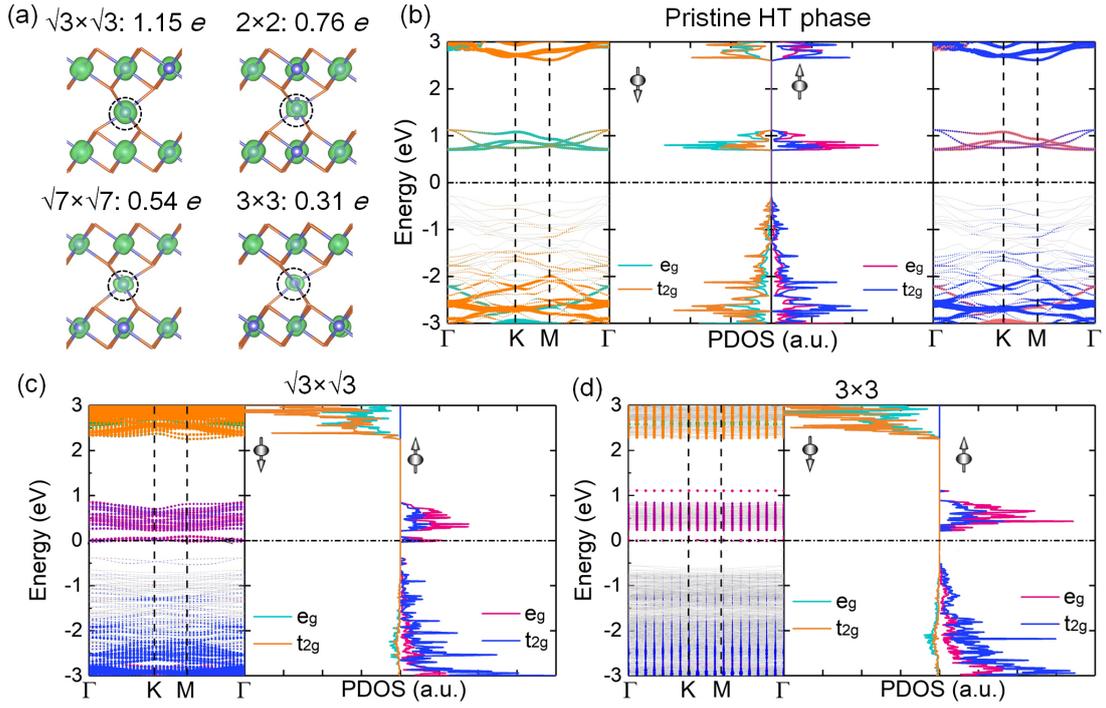

Fig. 3. (a) Decomposed spin-up charge density of conduction band minimum for Cr-intercalated bilayers in √3×√3, 2×2, √7×√7 and 3×3 supercell, respectively. The isosurface value is 0.05 e/Å³. The charge values at the intercalated Cr atom are present. (b-d) Band structures and project density of states (PDOS) of pristine HT phase of CrI$_3$ bilayer, Cr-intercalated HT phase in √3×√3 and 3×3 supercell, respectively.

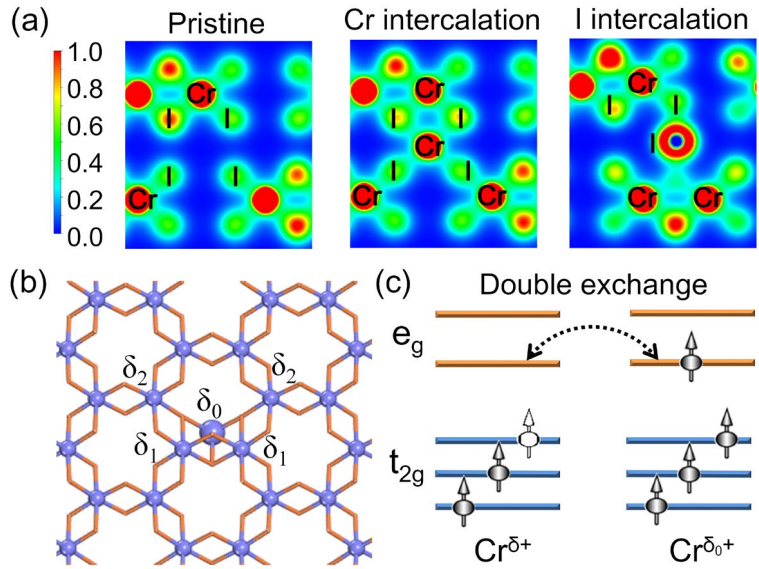

Fig. 4. (a) Contour plots for the electron densities of pristine, Cr-intercalated and I-intercalated HT phase of CrI$_3$ bilayer from left to right panels. (b) One layer of Cr-intercalated HT phase with the formal valence state $\delta_0$ (+2) for intercalated Cr atom, $\delta_1$ (+8/3) and $\delta_2$ (+17/6) for intralayer Cr atoms, and the valence state of other unmarked Cr atoms is +3, i.e., $2 = \delta_0 < \delta_1 < \delta_2 < 3$. (c) Schematic diagram of the double exchange mechanism for Cr/I-intercalated CrI$_3$ bilayer. The hollow arrow indicates that the electron is not fully occupied for the orbital.

# Supplementary Material

# Enhanced ferromagnetism of CrI$_3$ bilayer by self-intercalation

Yu Guo, Nanshu Liu, Yanyan Zhao, Xue Jiang, Si Zhou,[*] Jijun Zhao

*Key Laboratory of Materials Modification by Laser, Ion and Electron Beams, Dalian University of Technology, Ministry of Education, Dalian 116024, China*

[*] Corresponding author. Email: sizhou@dlut.edu.cn

**S1. Atomic structures of intercalated CrI$_3$ bilayers**

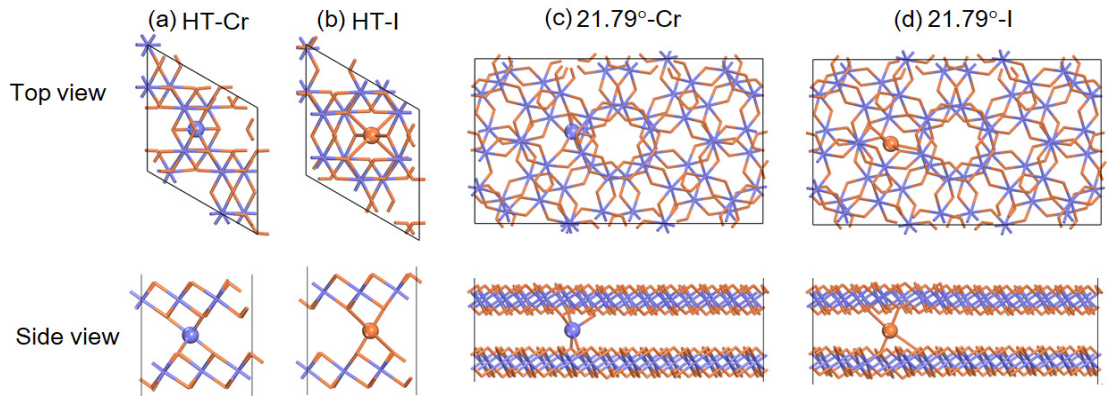

Fig. S1. Atomic structures of (a) Cr and (b) I intercalated HT phase of √3 × √3 supercell, and (c) Cr and (d) I intercalated twisted 21.79°-CrI$_3$. The Cr and I atoms are shown in purple and orange colors, respectively.

## S2. The intercalation energy for intercalated bilayers

Table S1. Intercalation energy ($E_{int}$) for Cr and I intercalated CrI$_3$ bilayers in the LT and HT phases using different supercells as well as twisted bilayers.

| | $E_{int}$ (eV) | √3×√3 | 2×2 | √7×√7 | 3×3 | 21.79° | 38.42° |
|---|---|---|---|---|---|---|---|
| Cr intercalation | LT | −4.01 | −4.04 | −4.11 | −3.97 | −4.23 | −3.97 |
| | HT | −4.29 | −4.30 | −4.35 | −4.39 | | |
| I intercalation | LT | −0.18 | −0.13 | −0.21 | −0.31 | −0.37 | −0.41 |
| | HT | −0.69 | −0.75 | −1.23 | −1.61 | | |

**S3. The interlayer distance for intercalated bilayer**

Table S2. Interlayer distance for Cr and I intercalated $CrI_3$ bilayers in the LT and HT phases using different supercells as well as twisted bilayers. The numbers in the brackets are the exchange energies for twisted 21.79°-$CrI_3$ and 38.42°-$CrI_3$ without intercalation.

|  | phase | pristine | √3×√3 | 2×2 | √7×√7 | 3×3 | 21.79° | 38.42° |
|---|---|---|---|---|---|---|---|---|
| Cr intercalation | LT | 3.54 | 3.38 | 3.40 | 3.39 | 3.34 | 3.48 (3.64) | 3.40 (3.60) |
| | HT | 3.54 | 3.34 | 3.36 | 3.37 | 3.37 | | |
| I intercalation | LT | 3.54 | 4.04 | 3.86 | 3.58 | 3.50 | 3.70 | 3.78 |
| | HT | 3.54 | 4.26 | 3.87 | 3.59 | 3.52 | | |

**S4. Thermodynamical stability of intercalated CrI₃ bilayers**

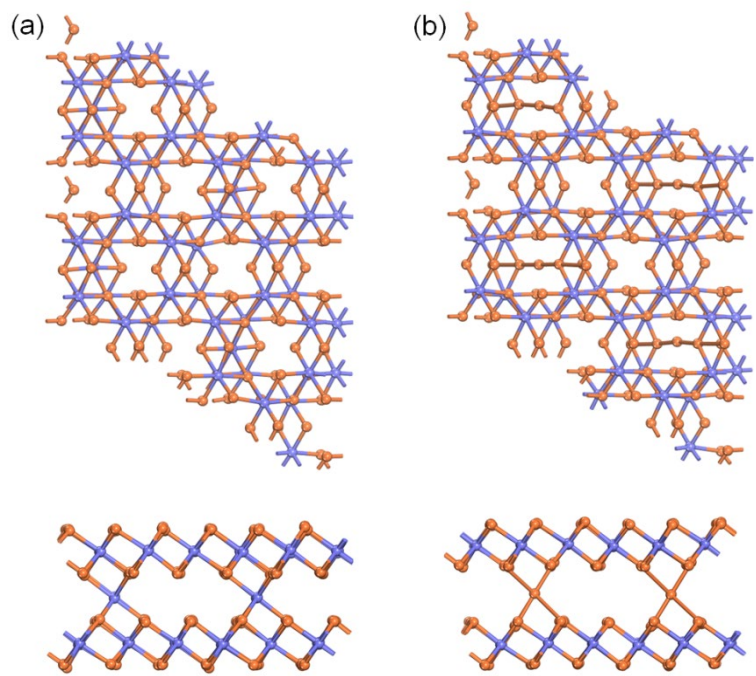

Fig. S2. Snapshots of (a) Cr and (b) I intercalated CrI$_3$ bilayers for HT phase from BOMD simulations with temperature controlled at 300 K. Each simulation is lasted for 10 ps.

**S5. Diffusion of intercalated Cr atom at van der Waals gap**

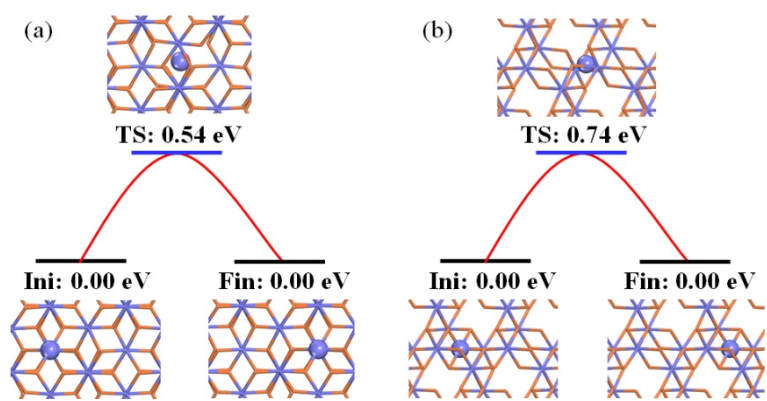

Fig. S3. Diffusion behavior of intercalated Cr atom in the van der Waals gap of (a) LT and (b) HT phases of bilayer $CrI_3$. The Cr and I atoms are shown in purple and orange colors, respectively. The intercalated Cr atoms are given in larger purple balls. The climbing-image nudged elastic band (CI-NEB) method was employed to investigate the diffusion kinetics and determine the activation energy for migration. Five images were used to calculate the diffusion path. The intermediate images of each CI-NEB simulation were relaxed until the perpendicular forces were smaller than 0.02 eV/Å.

## S6. Exfoliation behaviors of intercalated CrI$_3$ systems

We calculated the exfoliation energies of intercalated and pristine CrI$_3$ systems by simulating the separation of one CrI$_3$ layer from the intercalated and pristine bilayers. The equilibrium distance between intercalated bilayer and separated layer is 3.50 Å, which can be determined by the function of distance respect to total energy (Fig. S4a). Then we simulated the exfoliation process and predicted the exfoliation energy with respect to the separation distance, as shown in Fig. S4b. The calculated exfoliation energies are 0.21 J/m$^2$ for pristine systems and 0.20 J/m$^2$ for intercalated systems, respectively, indicating that the intercalated systems preserve the exfoliation behavior of pristine layered CrI$_3$.

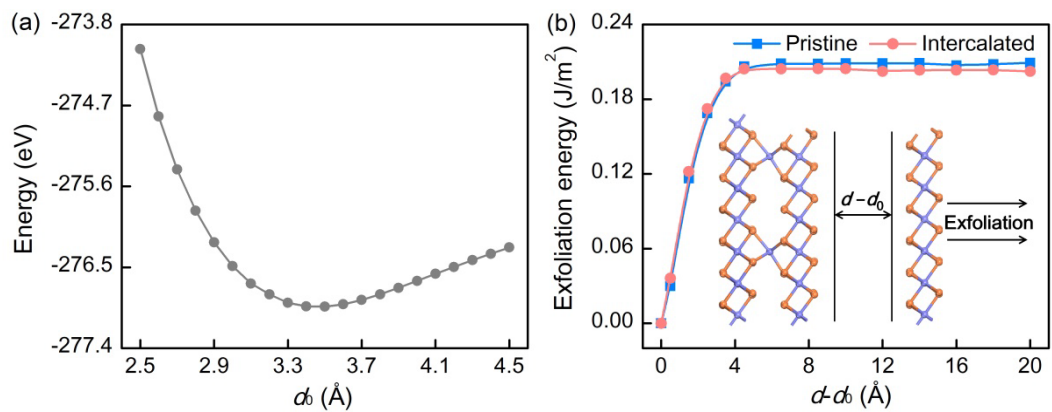

Fig. S4. (a) Total energy vs. the distance ($d_0$) between intercalated bilayer and a separated layer. (b) Exfoliation energy vs. separation distance $d$. for intercalated CrI$_3$ bilayer in comparison with pristine CrI$_3$ system..

**S7. Electronic structures of intercalated bilayers**

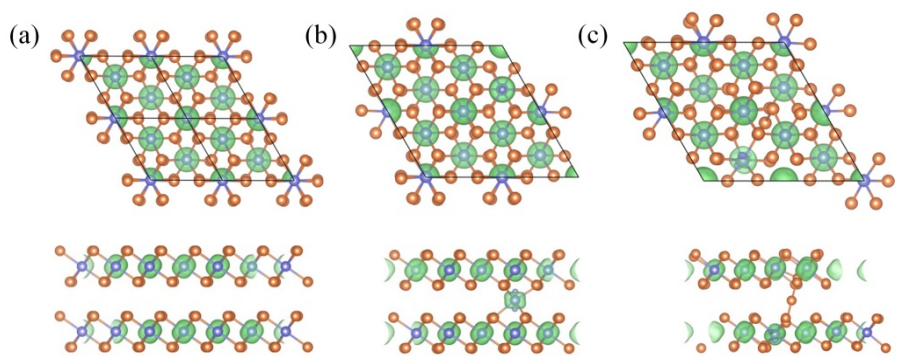

Fig. S5. Spin charge density (in green color) with an isosurface value of 0.01 e/Å$^3$ for (a) pristine, (b) Cr and (c) I intercalated CrI$_3$ bilayers, respectively. The Cr and I atoms are shown in purple and orange colors, respectively.

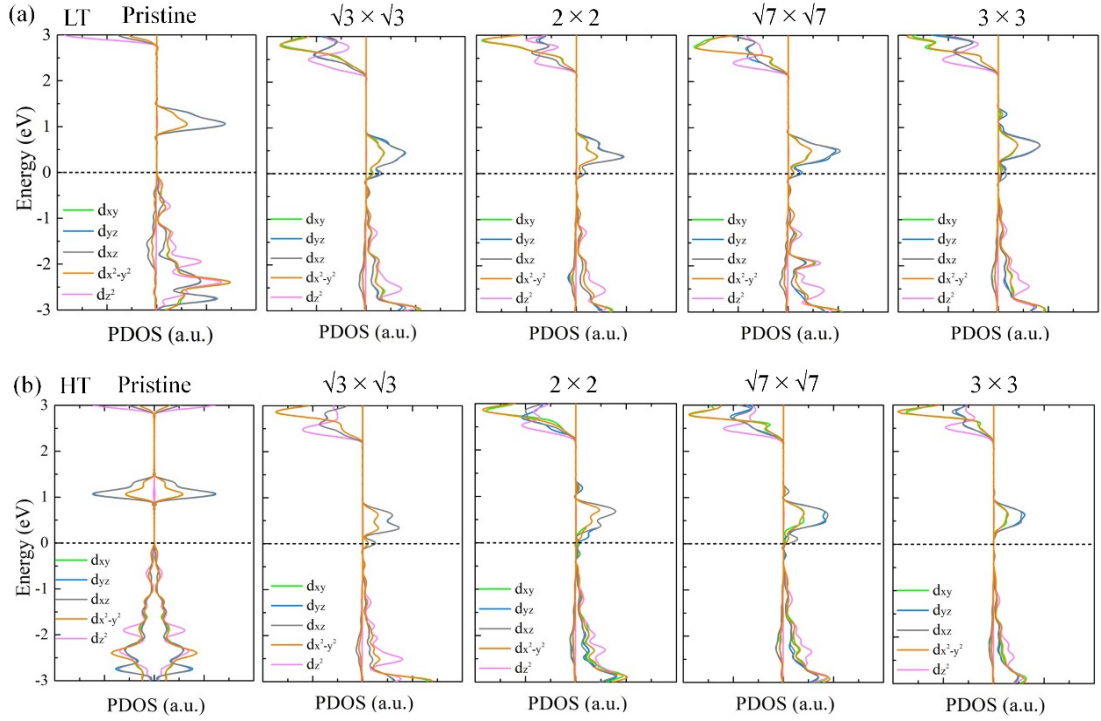

Fig. S6. Projected density of states (PDOS) of pristine and Cr-intercalated $CrI_3$ bilayers of (a) LT and (b) HT phases, with different supercells.

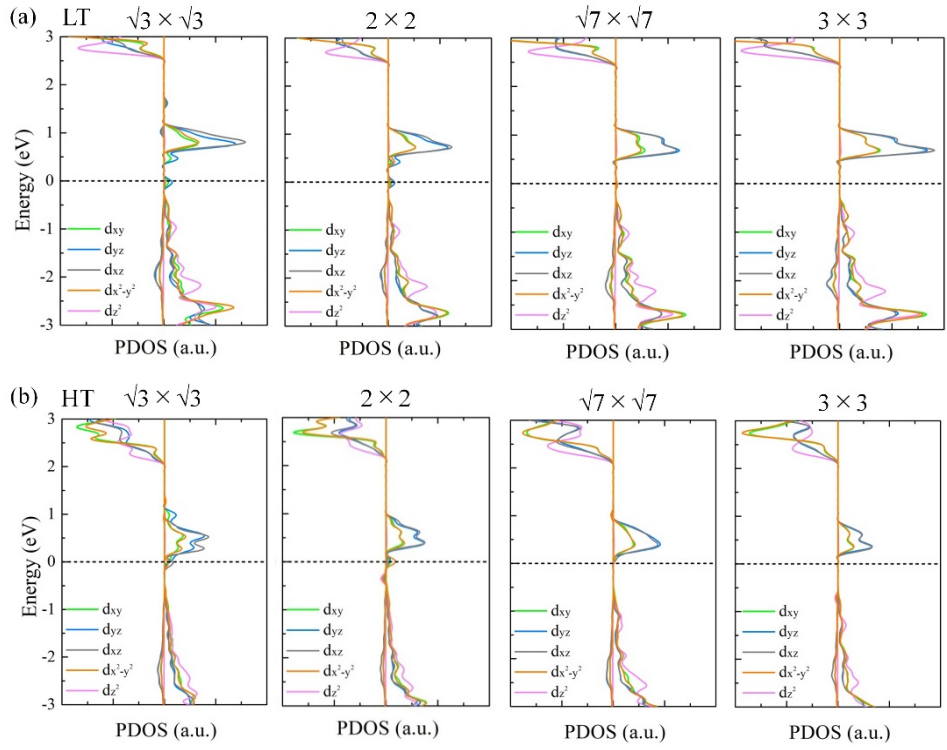

Fig. S7. Projected density of states (PDOS) of pristine and I-intercalated $CrI_3$ bilayers of (a) LT and (b) HT phases, with different supercells.

## S8. Self-intercalation for trilayer and bulk CrI$_3$

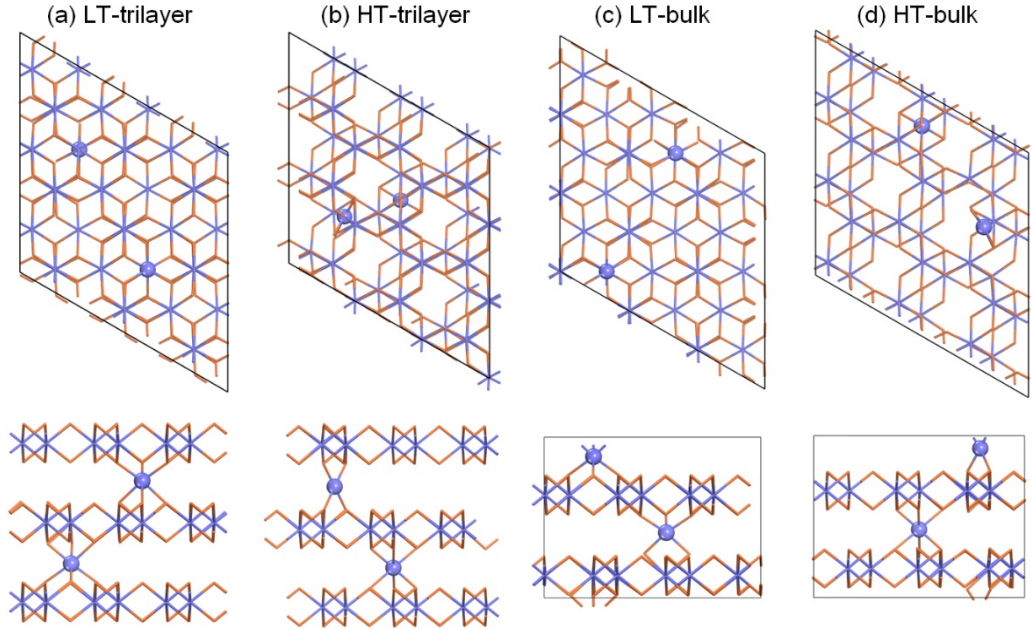

Fig. S8. Atomic structures of Cr intercalated (a) (c) LT and (b) (d) HT phases for trilayer and bulk CrI$_3$ in 3×3 supercell.

Table S3. Exchange energy $\Delta E$ of pristine, trilayer and bulk in LT and HT phases of 3×3 supercell. The atomic structures are shown in Fig. S8.

| $\Delta E$ (meV/f.u.) | Trilayer | | Bulk | |
|---|---|---|---|---|
| | Pristine | Intercalated | Pristine | Intercalated |
| LT | 2.03 | 7.01 | 3.30 | 19.66 |
| HT | −0.20 | 6.34 | −0.98 | 13.15 |

## S9. Virtual exchange gap for intercalated CrI$_3$ bilayers

Table S4. Virtual exchange gap ($G_{ex}$) of Cr-intercalated bilayer CrI$_3$ of the LT and HT phases with different supercells.

| $G_{ex}$ (eV) | pristine | √3×√3 | 2×2 | √7×√7 | 3×3 |
|---|---|---|---|---|---|
| LT | 0.98 | 0.37 | 0.44 | 0.74 | 0.83 |
| HT | 0.92 | 0.48 | 0.59 | 0.76 | 0.81 |

## S10. Charge transfer for the intercalated CrI$_3$ bilayers

Table S5. Charge transfer (CT, in the unit of electrons) of Cr and I-intercalated bilayer CrI$_3$ of the LT and HT phases with different supercells.

| | CT (*e*) | √3×√3 | 2×2 | √7×√7 | 3×3 |
|---|---|---|---|---|---|
| Cr intercalation | LT | 0.11 | 0.06 | 0.04 | 0.02 |
| | HT | 0.11 | 0.12 | 0.11 | 0.12 |
| I intercalation | LT | 0.08 | 0.06 | 0.07 | 0.06 |
| | HT | 0.10 | 0.06 | 0.11 | 0.11 |

**S11. Double exchange for I-intercalated bilayers**

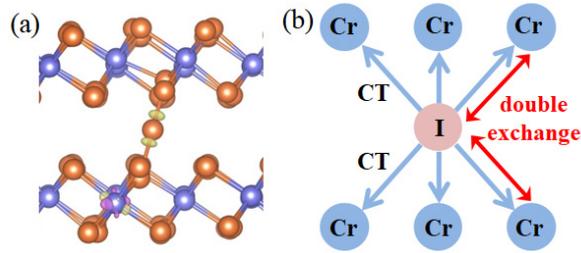

Fig. S9. Differential charge density of (a) I-intercalated LT phase of bilayer $CrI_3$. Yellow and pink colors represent the charge accumulation and depletion regions, respectively, with an isosurface value of $8\times10^{-3}$ $e/Å^3$. (b) Schematic illustrations of double exchange in the I-intercalated $CrI_3$ bilayer. Blue arrows show the charge transfer (CT) from the intercalated I atom to the intralayer Cr atoms, and red arrows highlight the double exchange interaction between Cr-Cr atoms.